\documentstyle[12pt]{article}

\begin{document}
\newcommand{\cl}{\centerline}
\newcommand{\beq}{\begin{equation}}
\newcommand{\eeq}{\end{equation}}
\newcommand{\beqa}{\begin{eqnarray}}
\newcommand{\eeqa}{\end{eqnarray}}
\def\d{{\rm d}}
\hbox{\hskip 4.0 true in ITP-US-93-8}
\vbox{\vskip 0.3 true in}
 \centerline{\large{{\bf Spectral Densities}}}\par
\centerline{\large{{\bf  and  Borel Transforms}}}\par
\centerline{\large{{\bf in Compton scattering}}}\par

\vbox {\vskip 0.3 true in}

\centerline{Claudio Corian\`{o}}
\centerline{ Institute for Theoretical Physics}
\centerline{ University of Stockholm}
\centerline{Box 6730, 113 85 Stockholm, Sweden}
\centerline{and}
\centerline{Institute For Theoretical Physics}
\centerline{State University of New York at Stony Brook}
\centerline{Stony Brook, 11794 NY, USA}
\vbox {\vskip 0.2 true in}

\centerline {ABSTRACT}
\medskip
\narrower{
 We show that the leading  double spectral density
in sum rules for  Compton-like  processes
can be obtained by simple properties of the Borel transform,
extending an approach  widely used in the literature on sum rules,
and known to be valid only for the spectral densities of form factors.
The extension is illustrated in the scalar case,
 where it is shown to be
consistent with Cutkosky rules.
Using  arguments based on the analiticity properties of the vertex and the box
diagrams,
we show that Compton scattering is, however, a favourable case and
indeed
possible disagreements between the two
methods are likely to be encountered in more general situations.


\newpage

\section{Introduction}
There are two main methods, developed in the past for 2 and 3 point functions
\cite{NR1,JS}
in QCD sum rules, which allow a calculation of the spectral function,
for any specific choice of the interpolating currents. One of them
is based on the cutting rules
for the propagator \cite{JS}, while a second method employs the Borel transform
in a rather original way \cite{NR1}.  While for 2-point function the
calculation of the
spectral density is rather straightforward and the cutting rules
are probably the simplest way to approach the problem, for 3-point functions
the calculations are not so obvious and an independent check of the result
obtained
in one way, by
some other  independent method, is most welcomed.  These checks have been done
in the
past  in the case of 3-point functions - to lowest order -  and the technical
value of these
results cannot be underestimated  since the analytic structure of vertex
functions,
for off-shell external legs, even to lowest order, is non trivial.
We remind  that  in the sum rule approach to exclusive reactions
the first contribution to the spectral density  for form factors comes from
a diagram which does not contain any gluon and the double discontinuity
 of it - the spectral density - is calculated from a simple triangle diagram.
When we consider 4-point functions, it is almost mandatory to check results
obtained
by one method (the cutting rules) with those obtained by the "Borel method" of
Ref. \cite{NR1} and show that indeed there is agreement between the two.
The example we have in mind is pion Compton scattering.
It has been discussed elsewhere \cite{CRS} that  this process can be
described, in the region of moderate $s,t,u$ invariants
by sum rule methods. The approach proposed is formulated in full agreement
 with the
usual canons of  QCD sum rules, but the larger complexity
of the method requires,  at various stages,  a consolidation of the methods
employed in
the derivation of the spectral density of this process and a clear
comprehension
of  its relation to the form factor case.
One of the most striking features of the leading spectral density for  Compton
scattering at fixed angle is the indipendence of such a function from the
virtualities
of the two pions, a fact which is completely new compared to the case of the
form
factors.  In this note we show that the agreement
between  the Borel method of Ref.
\cite{NR1} and the cutting rules of \cite{JS}
in the derivation of the hadronic spectral density
 can be understood from a rather  simple perspective, related to the
analiticity
of the spectral function  in the momentum transfer $t$.
Similar arguments can be formulated for pion Compton scattering.
A simple analytic continuation of the box diagram -in the case of Compton
scattering -
gives us the possibility to apply the Borel method to this more complex case
directly,
and gives a result which can be easily understood in terms of the Cutkosky
rules.
Our discussion here is of technical nature, we refer to  \cite{CRS}
for a more comprehensive treatment.

 \section{The Borel method}
Let's focus our attention on Fig. 1b.
This diagram gives the leading spectral function for form factors to lowest
order
in $\alpha_s$ \cite{JS}. The dashed lines describe the usual cutting rules for
the propagators.  If we assume scalar fermions in the loop, the spectral
density
 for such a diagram takes the form (with $p_1^2=s_1,\, p_2^2=s_2$)
\beq
\Delta_3 (s_1,s_2,t)=\int d^4 k\delta_+(k^2)\delta_+((p_1-k)^2)\delta_+((p_1-k
+ q_1)^2),
\label {trian}
\eeq
where the "+" in the delta functions characterize the momentum flow.
A direct calculation of this diagram gives
\beq
\label{delta3}
\Delta_3(s_1,s_2,t)={\pi\over  2 \left( (s_1 + s_2 -t)^2 - 4 s_2
s_2)\right)^{(1/2)}}.
\eeq
The evaluation can be easily carried out in the Breit frame of the two (pion)
lines
characterized by the virtualities $p_1^2$ and $p_2^2$.
It easy to recognize that (\ref{delta3}) does not develop any additional
singularity
for positive  $s_1,\ s_2$, for $t=(p_2 - p_1)^2$ in the physical region,
 which is the main reason why the same result can be obtained by
a completely different method \cite{NR1}.
Another important observation, in this context, is the independence of the
discontinuity
from any mass in the propagator (see also \cite{Eden}).
Since similar analiticity properties of
the spectral density turn out to be still valid for Compton scattering (in the
massless
case), we are going to illustrate in some more detail how  the  " Borel method"
\cite{NR1}  works first in the case of 3-point functions and, later on, for
Compton
scattering.
  In our rederivation of the spectral densities for the pion we hope
to clarify the connections between the cutting rules and
the  latter  approach.

We start from two-point functions.
  Let's consider the dispersion relation  of the polarization operator
\beq
\Pi(q_1^2)={1\over \pi}\int_{0}^{\infty} {
\,\Delta(s)ds\over (s-q_1^2)}
 \,\,\, + subtractions,
\label{pd}
\eeq
 with a singularity cut starting at $q_1^2=0$.
Eq. (\ref{pd}) can also be written in the form
\beqa
\Pi(q_1^2) &=&{1\over\pi}\int_{0}^{\infty}ds\int_{0}^{\infty}d\alpha
\,\Delta(s) e^{-\alpha(s-q_1^2)}, \,\,\,\,\ q_1^2<0,
\label{pd1}
\eeqa
where we have used the exponential parametrization of the
denominator.

The Borel transform in one variable is defined in its
differential version by the operator \cite{SVZ}
\beq
B(Q^2\to M^2)=lim_{\stackrel{Q^2,n\to \infty}{Q^2/n=M^2}}\frac{1}{(n-1)!}
(Q^2)^n (-\frac{d}{dQ^2})^n.
\label{dif}
\eeq
$M^2$ denotes the Borel mass.

It satisfies the identity
\beq
B(Q^2\to M^2) e^{-\alpha Q^2}=\delta(1-\alpha M^2) \,\,\,\,\ \alpha Q^2>0.
\label{boq}
\eeq
Acting on the polarization operator $\Pi(q_1^2)$
with the Borel transform we get the usual exponential suppression of the higher
states
\beq
M^2 B(-q_1^2\to M^2)\Pi(q_1^2)={1\over \pi}\int_{0}^{\infty}ds
\,\Delta(s)e^{-s/M^2}.
\label{MBP}
\eeq
At this point we can Borel transform once again eq.~(\ref{MBP}),
with respect to the inverse Borel mass $1/M^2$, in order to obtain
\beq
B(1/M^2\to \nu)( M^2 B(-q_1^2\to M^2)\Pi(q_1^2))=\frac{1}{\nu}\Delta(1/\nu).
\label{BMMP}
\eeq
Eq. (\ref{BMMP})  shows that by applying
in a sequence Borel transforms
on $\Pi(q_1^2)$ we obtain an expression from which we can
 easily identify the spectral weight $\Delta(s)$ of eq.~(\ref{pd}).
Let's now come to 3-point functions.
The Borel transformed amplitude for the pion form factor is
given by \cite{NR1}
\[
\phi(M_1^2,M_2^2,q^2)={1\over \pi^2}\int_{0}^{\infty}ds_1\int_{0}^{\infty}ds_2
\,\rho_{\pi 3}^{pert}(s_1,s_2,q^2)\, e^{-s_1/M_1^2 - s_2/M_2^2}
\]
\beqa
&& ={3\over 2 \pi^2 (M_1^2 + M_2^2)}\int_{0}^{1} dx\, x (1-x)
\,exp\left({-x q^2\over(1-x)(M_1^2 + M_2^2}\right).
\label{fi}
\eeqa
To isolate the spectral function from
 eq.~(\ref{fi}) we need to use Borel transforms
and act on it with the differential operator
\beq
\label{BO}
B(1/M_1^2\to 1/\nu_1)B(1/M_2^2\to 1/\nu_2) M_1^2 M_2^2.
\eeq
Here we discuss a possible way of doing this.
The inversion of (\ref{fi}) can be obtained by taking
inverse Lapace transforms twice
- respect to $1/M_i^2$ - of this equation.
In fact, for a given function $F(M^2)$, the following identity
\beq
 {L}^{-1}(1/M^2\to\nu)F(M^2)=(1/\nu) B(1/M^2\to 1/\nu)F(M^2)
\eeq
relates the differential operator given in (\ref{dif}) to the inverse Laplace
transform
\beq
 {L}^{-1}(1/M^2\to\nu)={1\over 2 \pi i}
\int_{c-i\infty}^{c+ i \infty}d(1/M^2)\, exp\left ({\nu/M^2} \right ).
\label{inv}
\eeq
Defining $1/M_1^2= \mu_1, 1/M_2^2 =\mu_2$, then , we are
required to act with the operator defined in eq.~(\ref{BO}) on the integral
function
\beq
\chi(\mu_1,\mu_2,q^2)={1\over \mu_1 + \mu_2} \int_{0}^{\infty}dx\, { x
\over (x+1)^4}\,exp\left(-x Q^2 \mu_2 + {x q^2\mu_2^2 \over \mu_1 +
\mu_2}\right).
\label{chi}
\eeq

Using the gaussian relation
\beq
exp\left(\alpha^2/(4 k)\right)=\left(k/\pi\right)^{1/2}\int_{-\infty}^{\infty}
d\sigma \,exp({-k
\sigma^2 - \alpha \sigma} )
\label{gau}
\eeq
we can rewrite eq. (\ref{chi}) into the form

\beqa
&& \chi(\mu_1,\mu_2,q^2)  \nonumber \\
&& =\int_{-\infty}^{\infty}
d\sigma \int_{0}^{\infty}  dx {x\,exp\left({-x q^2 \sigma - (\mu_1 +\mu_2)
\sigma^2 -2 x^{1/2} q \mu_2 \sigma}\right)
\over (x+1)^4(\pi\, (\mu_1 + \mu_2))^{1/2}}.
\label{chimu}
\eeqa
By using the relation
\beq
 {L}^{-1}(\mu_1\to\nu_1) {1\over(\mu_1 +\mu_2)^{1/2}}exp\left(-\mu_1 +\mu_2
\sigma^2\right)
=\left( {\pi\over(\nu_1-\sigma^2)}\right)^{1/2} \Theta(\nu_1-\sigma^2)
\label{inv1}
\eeq
on (\ref{chimu}) we finally get
\beqa
&& {L}^{-1}(\mu_2\to\nu_2) {L^{-1}}(\mu_1\to\nu_1) \chi(\mu_1,\mu_2,q^2)
\nonumber \\
&& =  {L^{-1}}(\mu_2,\nu_2) \int_{0}^{\infty}{x dx\over (x+1)^4}
\int_{-\sqrt{\nu_1}}^{\sqrt{\nu_1}} d\sigma
 {exp\left ( (-x q^2 \mu_2 -2 x^{1/2}q \mu_2 \sigma -\mu_2)
\nu_1\right )\over (\nu_1-\sigma^2)^{1/2}}  \nonumber \\
&&  \quad \quad \quad  = \int_{\sqrt{\nu_1}}^{\sqrt{\nu_1}}d\sigma{(-\sigma +
f^{1/2})
^3 q^4\over \left(\sigma - f^{1/2})^2 + q^2 \right)^4  f^{1/2} (\nu_1-
\sigma^2)^{1/2}}\nonumber  \\
&& \quad \quad \quad = (1/6) q^4 (d/dq^2)^3 Y(q^2,\nu_1,\nu_2),
\label{inv2}
\eeqa
where
\[
Y(q^2,\nu_1,\nu_2)=\int_{-\sqrt{\nu_1}}^{\sqrt{\nu_1}} d\sigma {(\sigma -
f^{1/2})^3\over f^{1/2}(\nu_1- \sigma^2)^{1/2} \left((\sigma -f^{1/2})^2 +
q^2\right)
}, \]

\beq
f=\sigma^2 - \nu_1 -\nu_2.
\label{Yq}
\eeq
By redefining $\nu_{i}=s_{i}, \, i=1,2$, and after doing the explicit
integration
 of in eq.~(\ref{Yq}),  it is easy to relate this last result
to the 3-particle cut integral for
the triangle diagram   (Fig. 1b), in the scalar case,
\beq
\label{Yoq}
{Y(q^2, s_1,s_2)\over 4 q^2} =\Delta_3 (s_1,s_2,t)
\eeq
 The derivative with respect to $q_1^2$ in eq. (\ref{inv2})
 takes into account
the fermionic character of the propagators in Fig. 1b compared to the scalar
case
(see eq. (\ref{Yoq})).
 The spectral function for the pion form factor
can be expressed in the form \cite{NR1}
\beqa
&&{\rho_\pi}^{pert}_3(s_1,s_2,t) \nonumber \\
&& \quad \quad =
{3\over 2 \pi^2} t^2 \biggl( \left ({d\over dt}\right )^2 +
{t\over 3} \left ({d\over dt}\right )^3 \biggr)
{1\over ((s_1 + s_2 - t)^2 - 4 s_1 s_2)^{1/2}}
\label{ropi}
\eeqa

with $t=-q^2$.

For four-point functions this procedure simplifies considerably. The method can
be applied
exactly as in the form factor case, although it is necessary to work in the
euclidean
region from the beginning.

To be specific let's consider the full contribution to the box diagram (not
just its cut)
related to Fig.~1 and, for simplicity,
let's restrict our considerations to the scalar case.
We consider therefore the following 4-point function
\beq
T_4=\int {d^4 k \over k^2 (p_1-k)^2 (p_1-k+q_1)^2 (p_2-k)^2}
\label{t4}
\eeq
with massless propagators.
At fixed angle and with
\beq
s+t+u=p_1^2 + p_2^2 \,\,\,\,\, s>0,\,\,t<0, \,\,u<0
\label{con}
\eeq
and moderately large $s,t,u$,
using Cauchy's theorem on $T_4$, we can assume the validity
of a spectral representation for such integral
of the form
\beq
\label{t4g}
T_4=-{1\over 4 \pi^2}\int_{\gamma 1} ds_1\int_{\gamma 2}ds_2
{\Delta (s_1,s_2,s,t)\over (s_1 - p_1^2)(s_2 - p_2^2)}
\eeq
where the contours $\gamma_{1,2}$ are again chosen as in Fig.~2.
If we introduce a double spectral function $\Delta(s_1,s_2,s,t)$
we can rewrite $T_4$ as
\beq
T_4(p_1^2,p_2^2,s,t)=-{1\over 4
\pi^2}\int_{0}^{\lambda^2}ds_1\int_{0}^{\lambda^2}ds_2
{\Delta(s_1,s_2,s,t)\over (s_1-p_1^2)(s_2-p_2)^2} + . . .
\label{t4la}
\eeq
where the neglected pieces involve a complex part of the contour.
As we've already discussed the usual definition of the transform given by
(\ref{dif})
does not apply
 to this case since the virtualities in (\ref{t4g}) $p_1^2$ and $p_2^2$ are
forced to stay
inside the area delimited by the contour in Fig.~4.
In Ref. \cite{CRS}, using Cutkosky rules,
it has been shown that the leading perturbative spectral function
can be obtained, for this diagram, within the conditions  on $s$ and $t$
given by (\ref{con}), by the 3-cut integral (see Fig.~1a)
\beq
J(p_1^2,p_2^2,s,t,m=0)=\int
d^4k{\delta_+(k^2)\delta_+((p_1-k)^2)\delta_+((p_2-k)^2)\over
(p_1-k+q_1)^2}.
\label{Jzero}
\eeq
The evaluation of (58) has also been discussed in \cite{CRS} and the answer,
for the associated spectral density, turns out to be rather
simple
 \beqa
\Delta(s_1,s_2,s,t)&=&(-2\pi i)^3 J(p_1^2,p_2^2,s,t) \nonumber \\
                               &=& -{4 i\pi^4\over s t}.
\label{set}
\eeqa
Compared to the form factor case (see eq. (\ref{delta3}))
the discontinuity
along the cut $\Delta$ as given by (\ref{set}) is
$s_1$ and $s_2$ independent.
However such a simplification comes from having
neglected the quark masses in eq. (\ref{t4}) and having not considered other
subleading cuts
in the evaluation of $\Delta(s_1,s_2,s,t)$, which are suppressed by power of
$s, t $ or $u$, compared to the leading result.
Here we intend to show that the result in
(\ref{set}) can be  reobtained by the "Borel method" discussed in the previous
section.
Let's first notice that the spectral density, if we allow massive scalar
propagators in eq. (\ref{t4}), even in the scalar case,
is rather different from the expression given by eq. (\ref{set}). Again, in
this case as in
the massless
one, the leading spectral function is in fact obtained from the 3-particle cut
integral
\beq
J(p_1^2,p_2^2,s,t,m)=\int d^4 k {\delta_+(k^2-m^2)\delta_+((p_1-k)^2-m^2)
\delta_+((p_2-k)^2 - m^2)\over (p_1-k+q_1)^2 - m^2}.
\label{jem}
\eeq
and it can be expressed in the form
\[
\Delta(s_1,s_2,s,t,m) = (-2 \pi i)^3 J(s_1,s_2,s,t,m)
\]
\beq
=-{4 \pi^4 i \over\sqrt{-t (4 m^2 s_1 s_2 - 4 m^2 s s_2 -4 m^2 s s_1 + 4 m^2
s^2 + 4 m^2 s t - s^2 t)}}
\label{djem}
\eeq
which reproduces eq. (\ref{set}) in the $m=0$ case. Compared to (\ref{set}),
eq. (\ref{djem}) shows that the spectral function developes a singularity at
$s_2$ dependent positions in the $s_1$ plane, as it is expected in general.
This shows that the simple result given by eq. (\ref{set}) is a due to the
approximation
of massless quarks that we have considered.
 However,
 given the fact that the leading discontinuity of $T_4$, for positive $p_1^2$
and
$p_2^2$, is constant
 at fixed angle, eq. (\ref{set}) gives us the indication that
we can extend the dispersion integral in (\ref{t4la}), with suitable
subtractions, to
infinity, without worrying abuot eventual additional thresholds which might be
encountered
for large positive virtualities of the pions in the dispersion relation.
This more general feature of the spectral density (its mass dependence), as we
mentioned
above, is new compared to the form factor case.
This also explain why both Cutkosky rules and the "Borel method", illustrated
above,
converge toward a unique answer in the case of vertex functions.

 \section{Application to Compton scattering}
In order to confirm our reasoning, let's show that
 the scalar $\Delta(s_1,s_2,s,t)$ given by eq (\ref{set}) can be reobtained by
the "Borel
method" described above, a method whose validity relies primarily on the
assumption that a
given amplitude can be described by a suitable spectral
representation. Once this fact is accepted, we act with Borel transforms
on (\ref{t4la}) with $\lambda^2 $ going to infinity, exactly as in the form
factor case.

Let's work in the euclidean region of $T_4(p_1^2,p_2^2,s,t)$
with
spacelike external invariants ($p_1^2,p_2^2<0$,
$s=(p_1+q_1)^2<0$, $t=(q_2-q_1)^2<0$).
It is then possible to relate $T_4$ to its euclidean continuation $T_{4E}$
and use the Schwinger parametrization for the latter

\beq
T_{4E}=\int d^4k\,\int_{0}^{\infty} [d\alpha_i]\,exp\left ({-\alpha k^2 - \beta
(p'_1-k)^2
-\gamma (p'_1-k+q'_1)^2 -\epsilon (p'_2-k)^2}\right ),
\label{ex}
\eeq
where $\alpha_i$ is a short notation for the proper time parameters.
Divergences, in this representation, reappear when we move into the physical
$s,t$ region.
We perform the integration over the loop momentum in eq. (\ref{ex}) to get
\beq
T_{4E}=\int_{0}^{1}dx_1dx_2 dx_3 dx_4\delta(1-x_1-x_2-x_3-x_4)
\int_{0}^{\infty}\Sigma^3 d\Sigma e^{\tau}
\label{ex1}
\eeq
where
\beq
\tau=-\Sigma(A_1(x_i) s_1 -A_2(x_i) s_2 - A_3(x_i,t))
\eeq
with suitable expression for $A_1,A_2,A_3$.
$T_{4E}$ is related to $T_4$
by the analytic continuation
\beq
T_4({p}_1^2,{p}_2^2,s,t)=i T_{4E}({p'}_1^2,{p'}_2^2,s',t')
\label{eu}
\eeq
in the region where $p_1^2=-{p'}_1^2$
$p_2^2=-{p'}_2^2$ and, in general, $p_i\cdot p_j=-{p'}_i\cdot {p'}_j$.
 As we have discussed before, we can send the radius $\lambda^2$ of
 eq. (\ref{t4la}) to infinity and introduce suitable subtractions in the
corresponding
 dispersion integral in order to get
\beq
T_4 =-{1\over 4 \pi^2}\int_{0}^{\infty}\int_{0}^{\infty}ds_1 ds_2
{\Delta (s_1,s_2,s,t)\over (s_1-p_1^2)(s_2-p_2^2)} \,\,\,+\,\,\, subtr.
\label{ext}
\eeq
which, in the euclidean region, becomes
\beq
T_{4E} =-{i\over 4 \pi^2}\int_{0}^{\infty}ds'_1\int_{0}^{\infty}ds'_2
{\Delta
(-s'_1,-s'_2,s',t')\over (s'_1 +{p'_1}^2)(s'_2 +{p'_2}^2)}\,\,\, +\,\,\,\,
subtr.
\label{exte}
\eeq
Exactly as in the case of the polarization operator (eqs. (\ref{pd}) and
(\ref{pd1})),
we can now apply Borel transforms on eq.~(\ref{exte}) to get
 \beqa
\eta_E(M_1^2,M_2^2,s',t')\nonumber \\
&&= B({p'_1}^2\to 1/M_1^2)\,B({p'_2}^2\to
1/M_2^2)\,T_{4E}({p'}_1^2,{p'}_2^2,s',t')  \nonumber \\
&=& -{i\over 4\pi^2}\int_{0}^{{\infty}}ds'_1\int_{0}^{{\infty}}ds'_2
e^{-s'_1/M_1^2
-s'_2/M_2^2} \Delta(-s'_1,-s'_2,s,t) \nonumber  \\
 &&= \pi^2\int_{0}^{1}dx_1\,dx_2\,dx_3\,\delta(1-A_1 \Sigma M_1^2)\delta(1-A_2
M_2^2\Sigma ) e^{-\Sigma A_3} \Sigma^3 \,d\Sigma\nonumber \\
&&=\pi^2 \int_{0}^{1}{ dx_1 \,dx_2 \pi^2\over M_1^2
M_2^2 \,x_1^2 \,x_2}e^{-b_1/M_1^2 -b_2/M_2^2} \\
\label{eta}
\eeqa
where we have defined
\beqa
b_1 &=& s {(1-x_1-x_2)\over x_2}\nonumber  \\
b_2 &=& {(-s x_1 + t x_2)\over x_1}.
\label{bi}
\eeqa
Notice that $B({p'_i}^2\to M_i^2)=B(-p_i^2\to M_i^2)$, since $p_i^2<0$
by assumption.

By acting similarly on $T_4$  as given by (\ref{ext})  - with $B(-p_i^2\to
1/M_i^2),$ -
we also get
\beqa
\eta (M_1^2,M_2^2,s,t) &=&-{1\over 4
\pi^2}\int_{0}^{\infty}ds_1\int_{0}^{\infty}ds_2\,
\Delta(s_1,s_2,s,t)e^{-s_1/M_1^2-s_2/M_2^2} \nonumber  \\
 &= & B(-p_1^2\to M_1^2)\,B(-p_2^2\to M_2^2)\,T_4 (p_1^2,p_2^2,s,t).
\label{eta1}
\eeqa

 Reapplying the Borel transform (this time with respect to $1/M_i^2$)
sequentially  on both $\eta$ and $\eta_E$ and using eq. (\ref{boq})
 we get respectively
\[
B(1/M_1^2\to \nu_1)\,B(1/M_2^2\to \nu_2)\,(M_1^2 M_2^2\,
\eta_E(M_1^2,M_2^2,s',t'))
\]
\beqa
 & =&\int_{0}^{1}\int_{0}^{1} {dx_1 dx_2\over x_1^2 x_2} \delta(1-b_1 \nu_1)
\delta(1-b_2\nu_2) \nonumber \\
 &=&-{1\over 4 \pi^2 \nu_1\nu_2}\Delta_E(-1/\nu_1,-1/\nu_2,s',t')
\label{uno}
\eeqa
and
\[
B(1/M_1^2\to\nu_1)\,B(1/M_2^2\to\nu_2)\,(M_1^2 M_2^2\,\eta(M_1^2,M_2^2,s,t))
\]
\beq
=-{1\over 4 \pi^2 \nu_1\nu_2}\Delta(1/\nu_1,1/\nu_2,s,t).
\label{due}
\eeq
As we can see from eq. (\ref{uno}), the integral on the two parameters $x_1$
and $x_2$
 is now trivial and can be performed straightforwardly.
A simple calculation then gives
\beq
\Delta_E(-1/\nu_1,-1/\nu_2,s't')={4\pi^4\over{\nu_1 \nu_2 s' t'}}.
\label{treE}
\eeq
Using the identification $1/\nu_1\to s_1, 1/\nu_2\to s_2$,
analytically continuing back with the prescriptions $s'\to -s$ $t'\to -t$
and using eq. (\ref{eu}) we finally get
\beq
\Delta(p_1^2,p_2^2,s,t)=-4i{\pi^4\over s t},
\eeq
which agrees with the one obtained from the cutting rules eq. (\ref{set}).
Within the Borel method difficulties related to the evaluation of discontinuity
integrals  are bypassed.
The application of the Borel  method to higher point functions, as we have
seen,
is simple, but, in principle, one should expect additional singularities
in the perturbative spectral functions which do not allow a dispersion relation
extending up to infinity in the plane of the two pion virtualities.

\section{Conclusions}
We have seen that there is a clear correspondence between two different methods
for the evaluation of the perturbative spectral functions,
both of them largely employed in the context of QCD sum rules. We have shown
that
their agreement can be extended to Compton-like processes and we have also
pointed out
some possible shortcomings.
In the massless case a comparison between the two approaches turns out to be
possible.

\vbox{\vskip 0.3 true in}
\centerline{Acknowledgements}
I thank Prof G. Sterman for suggesting this investigation and for discussions
and  Profs. A. Radyushkin  T. H. Hansson and H. Rubinstein for discussions.
This work was supported in part by th U. S. National Science Foundation, grant
PHY-9211367.

\newpage

\noindent
{\bf Figure Captions}
\bigskip

 \bigskip

 \noindent
 1a. Contribution to the spectral density for pion Compton scattering.
\smallskip

\noindent
1b.  The vertex diagram in the spectral density for  the pion form factor.
\smallskip

\noindent
2. The integration contours for the scalar amplitude in the $p_1^2$ plane,
which includes  possible  additional thresholds .
\smallskip

\end{document}